%% file: main.tex
\documentclass[10pt,conference]{IEEEtran}

\usepackage{hyperref}
\usepackage[sorting=none, maxbibnames=9, citestyle=numeric-comp, giveninits=true]{biblatex}

\usepackage{amsmath,amssymb,amsfonts}
\usepackage{algorithmic}
\usepackage{graphicx}
\usepackage{textcomp}
\usepackage{xcolor}
\usepackage{listings}
\usepackage[many]{tcolorbox} 
\tcbuselibrary{listings} 
\usepackage{float}
\usepackage{enumitem}
\usepackage{url}
\usepackage[nonumberlist,acronym,toc]{glossaries}

\def\BibTeX{{\rm B\kern-.05em{\sc i\kern-.025em b}\kern-.08em
    T\kern-.1667em\lower.7ex\hbox{E}\kern-.125emX}}

\definecolor{mygreen}{rgb}{0,0.6,0}
\definecolor{mygrey}{rgb}{0.9,0.9,0.9}
\definecolor{mycyan}{rgb}{0,0.7,0.7}
\definecolor{blue}{rgb}{0,0,1}

\bibliography{literatur}
\defbibheading{bibintoc}[\refname]{\section*{#1}
  \addcontentsline{toc}{section}{#1}
}

\lstdefinestyle{pythonstyle}{
    language=Python,
    numbers=left,
    tabsize=4,
    frame=single,
    rulecolor=\color{black},
    breaklines=true,
    showstringspaces=false,
    basicstyle=\ttfamily\footnotesize,
    keywordstyle=\color{blue}\bfseries,
    commentstyle=\color{mygreen},
    numbers=none               
}

\lstdefinestyle{cstyle}{
    language=C,
    tabsize=4,
    frame=single,
    rulecolor=\color{black},
    breaklines=true,           
    breakatwhitespace=false,   
    basicstyle=\ttfamily\footnotesize,  
    keywordstyle=\color{blue}\bfseries,
    commentstyle=\color{mygreen},
    showstringspaces=false,
    numbers=none               
}

\lstdefinestyle{mystyle}{
    keywordstyle=\color{blue}\bfseries,
    identifierstyle=\color{black},
    basicstyle=\ttfamily\footnotesize,
    commentstyle=\color{mygreen},
    stringstyle=\color{red},
    showstringspaces=false,
    morekeywords={function\_name, length\_parameters, context, sample\_parameters, connection\_table, application\_layer, hardware\_specifics\_of\_rag\_output, application\_layer\_details}
}

\newtcblisting{myboxes}[2][]{%
    colback=mygrey!5!white,
    colframe=mycyan!75!black,
    fonttitle=\bfseries,
    listing only,
    listing options={style=mystyle, breaklines=true}, 
    title={#2}, 
    boxsep=4pt, 
    left=2pt, 
    right=2pt, 
    top=2pt, 
    bottom=2pt, 
    fontupper=\footnotesize 
}

\makeglossaries

\begin{document}

\input{glossary}

\title{\fontsize{24pt}{28pt}\selectfont Automated Code Generation and Validation for Software Components of Microcontrollers}

\author{
\large Sebastian Haug\\
\footnotesize Munich University of Applied Sciences,\\ Munich, Germany\\
shaug@hm.edu
\and
\large Christoph Böhm\\
\footnotesize Munich University of Applied Sciences,\\ Munich, Germany\\
christoph.boehm@hm.edu
\and
\large Daniel Mayer\\
\footnotesize AGSOTEC GmbH,\\ Munich, Germany\\
daniel.mayer@agsotec.de
}
\maketitle

\begin{abstract}
This paper proposes a method for generating software components for embedded systems, integrating seamlessly into existing implementations without developer intervention. We demonstrate this by automatically generating hardware abstraction layer (HAL) code for GPIO operations on the STM32F407 microcontroller. Using Abstract Syntax Trees (AST) for code analysis and Retrieval-Augmented Generation (RAG) for component generation, our approach enables autonomous code completion for embedded applications.
\end{abstract}

\begin{IEEEkeywords}
Automated code generation, AST, RAG, Embedded systems
\end{IEEEkeywords}
\section{\textbf{Introduction}}

The demand for embedded software has surged with advancements in IoT and machine learning \cite{Justyna2023}. Automating code generation for embedded systems, especially for microcontrollers, is critical as it provides precise and contextually relevant software components for hardware-specific tasks. Traditional code generators accelerate development but are costly and complex to build. Tools like STM32CubeIDE provide a robust platform for embedded development by offering graphical configurations and automated initialization code generation. These tools focus on generating hardware setup and driver code while leaving application-specific logic to developers. This paper introduces an alternative approach that complements existing tools like STM32CubeIDE by focusing on autonomous generation of application-specific components. By analyzing code structures through AST and generating components using RAG, our proposed method provides an additional pathway for achieving hardware-specific code generation. This solution does not replace existing tools but expands the options available for embedded developers. 
The goal is to use RAG to seamlessly integrate into workflows alongside IDEs like STM32CubeIDE. This process is demonstrated by autonomously generating HAL code for GPIO operations on the STM32F407 microcontroller, bridging application and hardware layers.

\subsection{Outline}

This paper begins with the \textbf{\hyperref[sec:background]{background}} section, which introduces the foundational concepts and technologies, including AST and RAG, that underpin our approach. Subsequently, the \textbf{\hyperref[sec:implementation]{implementation}} section describes the methodology and technical aspects of the proposed solution. Following this, in the \textbf{\hyperref[sec:evaluation]{evaluation}} section, the experimental results are presented. In the \textbf{\hyperref[sec:discussion]{discussion}} section, we reflect on the broader implications of the findings and discuss the practical applicability of the method. The \textbf{\hyperref[sec:validity]{threats to validity}} section addresses potential limitations and challenges that could affect the validity of the results. To situate our work in the broader context, the \textbf{\hyperref[sec:related]{related work}} section reviews existing literature in the field. Finally, the paper concludes with the \textbf{\hyperref[sec:outlook]{outlook}} section, offering future directions and exploring the broader implications of this research.

\section{\textbf{Background}}
\label{sec:background}

\subsection{Abstract Syntax Tree (AST)}

An Abstract Syntax Tree (AST) is a tree-like representation of the structure of source code, serving as a fundamental component in many programming tools, including compilers, interpreters, and automated code generation systems \cite{aho2006compilers}. Each node in an AST represents a construct occurring in the code, such as expressions, statements, or declarations, organized hierarchically to reflect their relationships and dependencies.

\subsection{Hardware Abstraction Layer (HAL)}

The Hardware Abstraction Layer (HAL)\footnote{\gls{HAL}} provides a standardized interface for interacting with hardware, abstracting away hardware-specific details and complexities \cite{beningo2016apisvshals}. This layer allows higher-level software to communicate seamlessly with hardware components without requiring intimate knowledge of the underlying hardware. The HAL thus promotes code portability and reuse across different hardware platforms by offering a uniform interface for common hardware operations.

In Figure \ref{fig:ast_implementation}, we provide a visualization of the simplified code structure generated by our approach. This visualization highlights the distinct layers in our implementation: 

\begin{figure}[H]
    \centering
    \includegraphics[width=0.9\linewidth]{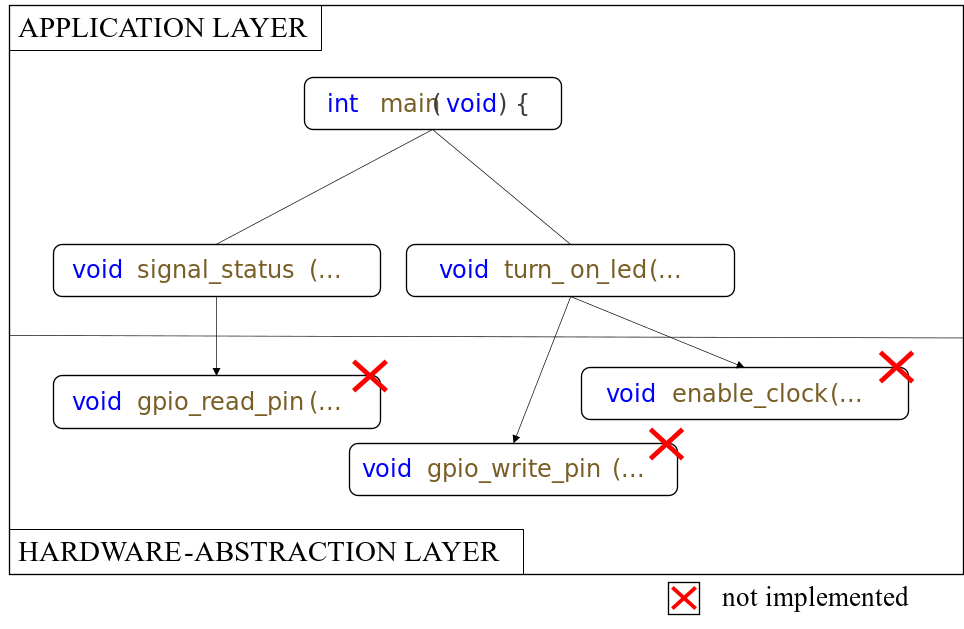}
    \caption[Functionality to Be Tested]{\textbf{Abstract Syntax Tree of software layers.} Visualization of the layers involved in code generation, showcasing the Application Layer and Hardware Abstraction Layer (HAL) interactions.}
    \label{fig:ast_implementation}
\end{figure}

At the top of the hierarchy is the \textbf{Application Layer}, where the main entry point of the program is defined, e.g., \texttt{int main(void)}.

Below the application layer is the \textbf{Hardware Abstraction Layer (HAL)}. The application layer interacts with the HAL by calling these hardware-specific functions. For example, to enable the clock for GPIO, read the pin state, or write to a GPIO pin, we generate the necessary HAL functions like \texttt{enable\_gpioa\_clk()}, \texttt{hal\_gpio\_read()}, and \texttt{hal\_gpio\_write()}.

This clear separation of layers allows us to abstract away hardware details while maintaining efficient control over low-level operations.

\subsection{Embeddings and Vector Stores}

A significant challenge in generating hardware abstraction layer (HAL) code is not only creating syntactically correct code but also ensuring that the generated code fits seamlessly into the larger codebase where it will be applied. To solve this, we use a LLM to generate code and vector stores to retrieve relevant contextual usage examples, ensuring that the generated HAL code integrates accurately and efficiently with existing systems.

A key component of our RAG-based approach is the use of embeddings and vector stores to provide contextually relevant information during code generation. Embeddings represent both language and code as dense vectors in a high-dimensional space, capturing semantic relationships that enable the model to understand and retrieve relevant code snippets. This structured representation allows the model to generate code that not only fits syntactically but is also contextually appropriate for the existing project.

To manage and search through large volumes of embeddings, we use FAISS (Facebook AI Similarity Search) \cite{douze2024faisslibrary}, an efficient tool for indexing and retrieving high-dimensional vectors. FAISS allows for real-time retrieval of similar code examples, guiding the generation of HAL code that aligns with the broader system, improving accuracy and seamless integration with the existing codebase.

\section{\textbf{Prototypical Implementation}}
\label{sec:implementation}

To validate the methods proposed in this paper, we developed a prototypical implementation using the Python programming language. 

\subsection{Microcontroller Selection}
\label{sec:microcontroller_section}
We primarily focused on the STM32F407\footnote{\gls{STM32F407}} microcontroller, which is widely used in various embedded systems due to its robust features and broad support in the developer community \cite{STM32F4DISCOVERY}.

\subsection{Integration of LLM-Model}
We integrate the GPT-4o Mini model\footnote{\gls{GPT4oMini}} into our system due to its affordability and sufficient capabilities for iterative testing of our techniques. The model serves as the core for generating code snippets based on provided prompts and context. By using the LangChain Python library, we manage API calls to the GPT-4o Mini model and utilize FAISS methods for efficient retrieval and similarity search, which enhances the context-awareness of the generated code.

To amplify the consistency of generated outputs, we set the model's temperature to 0 during experimentation. This configuration minimizes randomness in the responses, thereby reducing output variance across iterations. While this approach sacrifices some degree of creative diversity, it ensures that the generated outputs remain consistent and predictable. This consistency is a critical requirement for reliable code generation and validation in embedded systems.

Prompt Engineering is essential for leveraging Large Language Models (LLMs) to automate the generation of software components in microcontroller development. As described by Microsoft Azure, improving the design of prompts enables developers to effectively utilize LLMs to produce efficient, reliable, and maintainable code tailored to the specific requirements of embedded systems \cite{Urban2024}.

A structured prompt used in this paper includes several key components:

\subsubsection{Cue} The cue defines the function or identity that the automated system or tool will assume while executing the prompt. It sets the expectation and scope of the tasks that the system will handle, which is crucial in effective prompt design \cite{Urban2024}.

\begin{myboxes}[label=Cue Example]{} You will be my Custom Hardware Abstraction Layer Generator. \end{myboxes}

\subsubsection{Clear Instructions} Starting with clear instructions is vital, as it sets the foundation for the task at hand. Microsoft Azure advises that the sequence of information in the prompt significantly affects the model's response. Clear instructions include both a clear goal and constraints \cite{Urban2024}.

\begin{myboxes}[label={example_code}]{}
Please generate a custom C function implementation for the function '{function_name}' with {length_parameters} parameters like: {sample_parameters}.
\end{myboxes}

\subsubsection{Constraints}
Constraints outline the limitations or specific conditions that must be followed while fulfilling the prompt. They ensure that the generated output adheres to predefined rules and integrates seamlessly into the target system.

\begin{myboxes}[label={example_constraints}]{}
Don'ts:
- Don't reference new variables or functions that are not implemented.
- Don't reference stm32fxxx_hal.h functions.
...
\end{myboxes}

\subsubsection{Return Format}
This specifies how the results of the prompt should be structured and presented, enhancing the usability and clarity of the generated content \cite{deepset2024}.

\begin{myboxes}[label={example_return_format}]{}
Return-Format:
...
- Be well-documented with comments explaining its purpose, parameters, and return value.
- Create your own custom HAL functions without referencing other functions.
...
\end{myboxes}

\subsubsection{Supporting Context}
Context includes any background information, specific scenarios, or data that needs to be considered while executing the prompt. This helps tailor the response to fit the specific needs or circumstances of the project \cite{Urban2024}.

\begin{myboxes}[label={example_context}]{}
Create the {function_name} using the provided information about the existing code for an STM32F407 board: {context}
\end{myboxes}

\subsection{Automated Detection and Generation of Missing Code Elements}

Our approach enables a Python program to autonomously build upon an existing codebase by detecting and filling in missing code elements necessary for a complete Hardware Abstraction Layer (HAL). To achieve this, we employ an Abstract Syntax Tree (AST) representation of the program code, allowing the system to analyze code structure, relationships between functions, variables, and dependencies effectively \cite{aho2006compilers}. Using the AST, we can automatically identify missing components within the codebase and apply Retrieval-Augmented Generation (RAG) to generate the necessary functions and structures contextually.

\begin{figure}[H]
    \centering
    \includegraphics[width=0.9\linewidth]{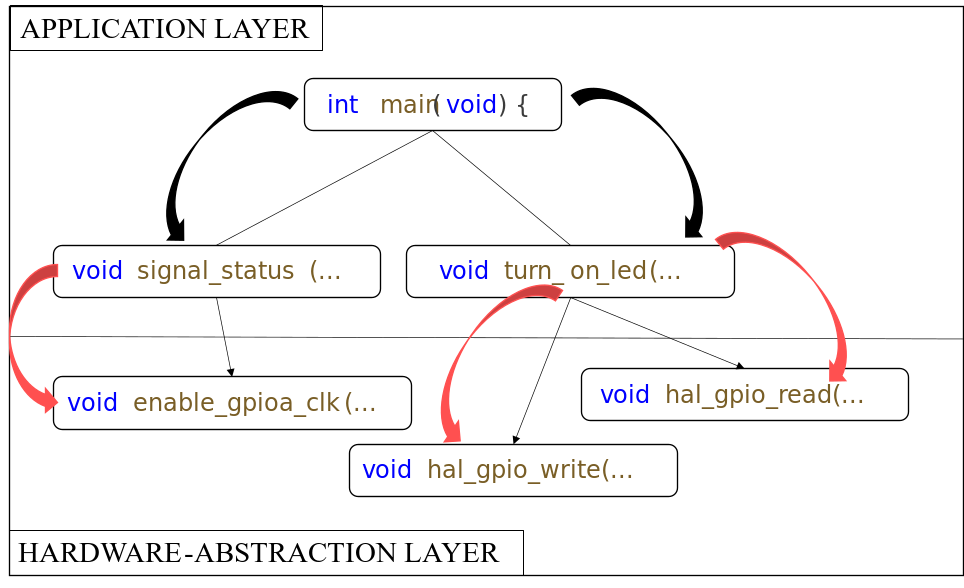}
    \caption[AST Implementation Process]{\textbf{AST Implementation Process.} This figure illustrates the process of using an Abstract Syntax Tree (AST) to identify and implement missing functions within a codebase.}
    \label{fig:ast_implementation_process}
\end{figure}

The process is visually represented in Figure \ref{fig:ast_implementation_process}, where the AST identifies missing low-level functions in the Hardware-Abstraction Layer that are required for the Application Layer to operate correctly. Functions like \texttt{gpio\_read\_pin} and \texttt{enable\_clock} are detected as missing and are marked for generation, ensuring that the application can interact seamlessly with hardware components.

For the prototype, the `pycparser` library was utilized to generate the AST, which serves as a basis for detecting missing functions and seamlessly integrating new code. This library facilitates an iterative process of code analysis and modification, minimizing the need for manual intervention in HAL generation.

\begin{figure}[H]
    \centering
    \includegraphics[width=1\linewidth]{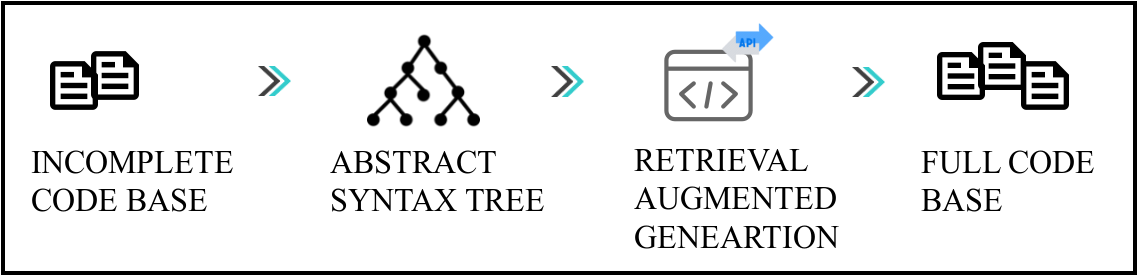}
    \caption[Code Generation Process]{\textbf{Code Generation Process.} Starting with an incomplete code base, the AST is traversed to analyze the structure and identify missing pieces. Retrieval-Augmented Generation (RAG) is then used to generate each missing segment, resulting in a fully complete code base.}
    \label{fig:code_generation_process}
\end{figure}

The overall workflow, from AST-based analysis to RAG-based generation, is depicted in Figure \ref{fig:code_generation_process}. This process starts by analyzing the program structure with the AST to locate gaps, then applies RAG to generate each missing element. 

In our implementation, we automatically generate hardware-specific functions optimized for GPIO\footnote{\gls{GPIO}} operations, such as:
\begin{itemize}
    \item \texttt{enable\_gpioa\_clk()} – Enables the clock\footnote{\gls{Clock}} for GPIOA.
    \item \texttt{set\_io\_mode()} – Configures a GPIO pin as input or output.
    \item \texttt{hal\_gpio\_write()} – Writes to a GPIO pin.
    \item \texttt{hal\_gpio\_read()} – Reads a GPIO pin state.
    \item \texttt{hal\_gpio\_toggle()} – Toggles a GPIO pin state.
\end{itemize}

In addition to functions, we automatically generate hardware-specific variables that define key offsets and base addresses for peripheral registers. Examples include: 
\begin{itemize} 
    \item \texttt{USART\_SR\_OFFSET} – Offset\footnote{\gls{Offset}} for the USART Status Register. 
    \item \texttt{USART2\_BASE} – Base address for the USART2 peripheral. 
    \item \texttt{USART\_FLAG\_TXE} – Flag indicating the transmit data register is empty. 
    \item \texttt{USART\_DR\_OFFSET} – Offset for the USART Data Register. 
    \item \texttt{RCC\_AHB1ENR\_OFFSET} – Offset for the AHB1 clock enable register. 
    \item \texttt{RCC\_BASE} – Base address for the Reset and Clock Control peripheral. 
    \item \texttt{USART\_BRR\_OFFSET} – Offset for the Baud Rate Register. 
    \item \texttt{USART\_CR1\_OFFSET} – Offset for Control Register 1. 
\end{itemize}

The result is a complete, functional codebase where all necessary HAL components are generated automatically, ensuring smooth integration with existing code.

\subsection{Automated Validation of Generated Code}

One of the primary challenges in validating the generated code is ensuring that it not only compiles but also functions as intended in its application context. To address this, we implemented a self-testing validation process. The validation begins with a compilation check to ensure that the generated code integrates without errors. Once the program successfully compiles, a series of automated tests is executed to verify that the program produces the correct outputs based on predefined test scenarios.

In this setup, the HAL (Hardware Abstraction Layer) logic is dynamically generated, while the application layer provides the program's intended functionality and expected test results. This separation enables a robust validation process where each layer contributes to ensuring functionality and correctness. By automating both compilation and functional testing, this approach minimizes the need for manual intervention, allowing for an efficient and reliable validation workflow.

For functionality validation, we utilized Renode, an open-source simulation framework designed for hardware simulation and testing \cite{RenodeFramework}. Renode provides the ability to simulate the necessary hardware addresses and peripherals required for our use case, such as the STM32F407 microcontroller's GPIO, USART, and clock registers. By using Renode, we were able to emulate the behavior of the microcontroller without requiring physical hardware, enabling a more flexible and efficient testing environment.

\begin{figure}[H]
    \centering
    \includegraphics[width=1\linewidth]{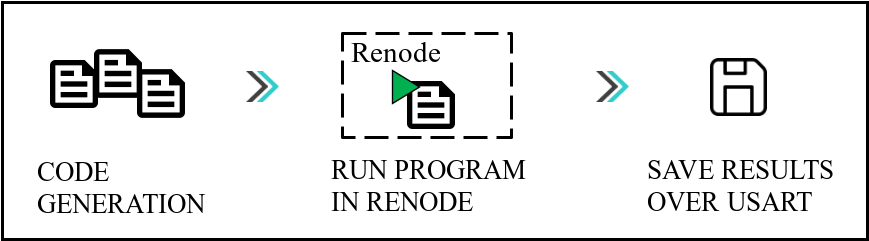}
    \caption[Code Validation Process]{\textbf{Code Validation Process.} After generating the code, the build version is run in a Renode environment. The application logs are sent over USART and saved for further analysis.}
    \label{fig:validation}
\end{figure}

After generating the code, the validation process follows the steps illustrated in Figure \ref{fig:validation}. This diagram details the validation workflow, including the automated testing environment and log analysis.

\section{\textbf{Experimental Evaluation}}
\label{sec:evaluation}

The experimental evaluation aimed to assess the feasibility and reliability of the proposed method for generating hardware abstraction layer (HAL) code that integrates seamlessly with an existing STM32F407 microcontroller project. This evaluation tested the accuracy, stability, and functional correctness of the generated code across multiple iterations and scenarios.

\subsection{Accurate Code Generation for STM32F407}

Our method consistently generated accurate and functional HAL code tailored to the STM32F407 microcontroller. One of the key functions, \texttt{set\_io\_mode()}, is shown below. This function configures the input/output mode of a GPIO pin by directly manipulating the microcontroller's registers:

\begin{lstlisting}[style=cstyle, caption={ \texttt{set\_io\_mode()} Function}]
void set_io_mode(uint32_t gpio_base, uint32_t pin_mask, uint8_t mode) {
    volatile uint32_t *GPIO_MODER = (uint32_t *)(gpio_base + 0x00);
    uint8_t pin_number = 0;
    while ((pin_mask >> pin_number) != 1) {
        pin_number++;
    }
    *GPIO_MODER &= ~(0x3 << (pin_number * 2));
    *GPIO_MODER |= (mode << (pin_number * 2));
}
\end{lstlisting}

This example highlights capability of the process to create hardware-specific code that adheres to the STM32F407's architecture and register layout.

The complexity of configuring GPIO ports and understanding the underlying hardware specifics highlights the challenges involved in embedded system engineering \cite{zhu2015embedded}. Such configurations require precise control and thorough knowledge of both the hardware and its interaction with software.

\subsection{Test Cases for Code Regeneration}

To evaluate the robustness of the code generation process, two distinct test cases were designed, simulating different levels of complexity in the regeneration process:

\subsubsection{Random Element Deletion and Regeneration}
In this scenario, a single random element (e.g., a function or variable) was deleted from the HAL, and the generator was tasked with regenerating it. Each iteration of this test case required a single API call to generate the missing element. This test measured the generator's ability to restore missing functionality while maintaining integration with the rest of the codebase.

\subsubsection{Complete HAL Deletion and Regeneration}
In the second test case, the entire HAL was deleted, requiring the system to regenerate all HAL components from scratch. This process involved 12 API calls for each iteration, as multiple functions, variables, and structures needed to be recreated. This test simulated the generator's performance in handling larger-scale regeneration tasks and validated its ability to reconstruct a cohesive and functional HAL.

\subsection{Building the Project with CMake}

The generated code was integrated into the existing project structure and compiled using CMake. This step ensured that the code adhered to the project's build requirements and successfully passed the compilation stage, verifying the syntactic correctness and completeness of the generated code.

\subsection{Hardware Emulation and Software Testing with Renode}

Renode enabled us to emulate the STM32F407 microcontroller and execute the software in a controlled virtual environment. This approach provided several advantages:
\begin{itemize}
    \item \textbf{Hardware Independence}: Testing could be performed without physical hardware.
    \item \textbf{Repeatability}: The simulated environment ensured consistent and reproducible test conditions.
    \item \textbf{Efficiency}: Functional validation could be conducted rapidly, bypassing the time-consuming process of deploying and testing on actual hardware.
\end{itemize}

By running the generated software in Renode, we were able to verify its functional correctness through application-layer tests integrated into the system. The application logs were transmitted over USART and analyzed to confirm expected behavior.

\subsection{Iterations and Results}

Each test case was executed for more than 100 iterations. Key observations include:
\begin{itemize}
    \item The process consistently produced syntactically correct and functionally accurate code.
    \item The correct functionality was sufficient across all iterations, even as structural variance was observed in the regenerated code.
    \item The regeneration process exhibited robust adaptability to varying levels of complexity, from individual elements to complete subsystems.
\end{itemize}

\section{\textbf{Discussion}}
\label{sec:discussion}

The experimental results demonstrate the feasibility and potential of autonomous code generation in embedded systems. Below, we discuss the key results and their implications.

\subsection{Feasibility of Autonomous Code Integration}

The results confirm that it is consistently possible to generate HAL code that integrates seamlessly into an existing codebase. This finding is significant, as it demonstrates the capability of autonomous generation systems to produce code that fits contextually without requiring extensive direction or refinement. This opens up the possibility of automating larger-scale coding projects across diverse environments and use cases, reducing the manual effort involved in writing low-level embedded code.

The ability to create working HAL code tailored to the STM32F407 microcontroller also suggests the potential for extending this approach to other microcontroller families. By adapting the generation process to specific hardware architectures, the system could serve as a versatile tool for embedded software development.

\subsection{Efficient Validation via Embedded Tests}

Embedding tests within the application layer proved to be a critical aspect of the validation process. This approach allowed us to verify the functionality of the generated software "on the run" in a simulated environment, bypassing the need for physical hardware testing. Given the engineering challenges and time-intensive nature of microcontroller testing, this method significantly streamlined the validation process.

By leveraging Renode for hardware emulation, we were able to test the generated software efficiently and reproducibly. The simulated environment ensured that tests could be conducted under controlled conditions, providing valuable insights into the behavior of the code without the variability introduced by physical hardware.

\section{\textbf{Threats to Validity}}
\label{sec:validity}

Since the experiment was conducted using a specific language model (ChatGPT-4-turbo) and particular code generation tasks involving HAL deletions, the results may not necessarily apply to different models or domains. Variations in the model architecture, training data, or domain-specific requirements may impact the behavior of generated code and the observed similarity trends. Additionally, while this study uses a single type of microcontroller code, it remains uncertain how well these results would transfer to other programming languages or embedded systems.

Another consideration is the rapid evolution of language models, including potential updates to the ChatGPT-4 or Mini API. As these models are refined, adjustments may need to be made to maintain the reliability and consistency of generated outputs. Future updates to the model could improve or alter its code generation capabilities, leading to variations in the prototype’s behavior. Therefore, while the underlying concept and methodology demonstrated in this study provide a proof of concept, the practical effectiveness of the approach may vary as the technology advances.

\section{\textbf{Related Work}}
\label{sec:related}

Recent approaches like the "Outline, Then Details" framework show promise in syntactically guided code generation for embedded environments \cite{OutlineThenDetails}.

Inspired by the self-planning approach in code generation\cite{SelfPlanningCodeGeneration}, this method applies structured decomposition, utilizing Abstract Syntax Trees (AST) to systematically analyze existing code and Retrieval-Augmented Generation (RAG) to generate and integrate HAL components seamlessly with minimal developer intervention. Following a phased methodology similar to self-planning, this approach leverages AST for intent analysis and RAG for targeted code retrieval, ensuring that generated components align with existing code structures, enhancing both integration quality and code maintainability.

Techniques like Retrieval-Augmented Code Generation (RAG) are becoming more prevalent, as they can efficiently generate missing software components by retrieving relevant examples from existing codebases \cite{RetrievalAugmentedCodeGeneration}. 

Retrieval-Augmented Generation (RAG) is a technique that integrates existing code components into new code generation by retrieving relevant information from a vector database. We employ RAG to enable efficient code completion, particularly for embedding pre-existing function blocks into hardware abstraction layer (HAL) code for seamless integration \cite{LLMRAGCodeGeneration}.

\begin{figure}[H]
\centering
\includegraphics[width=1\linewidth]{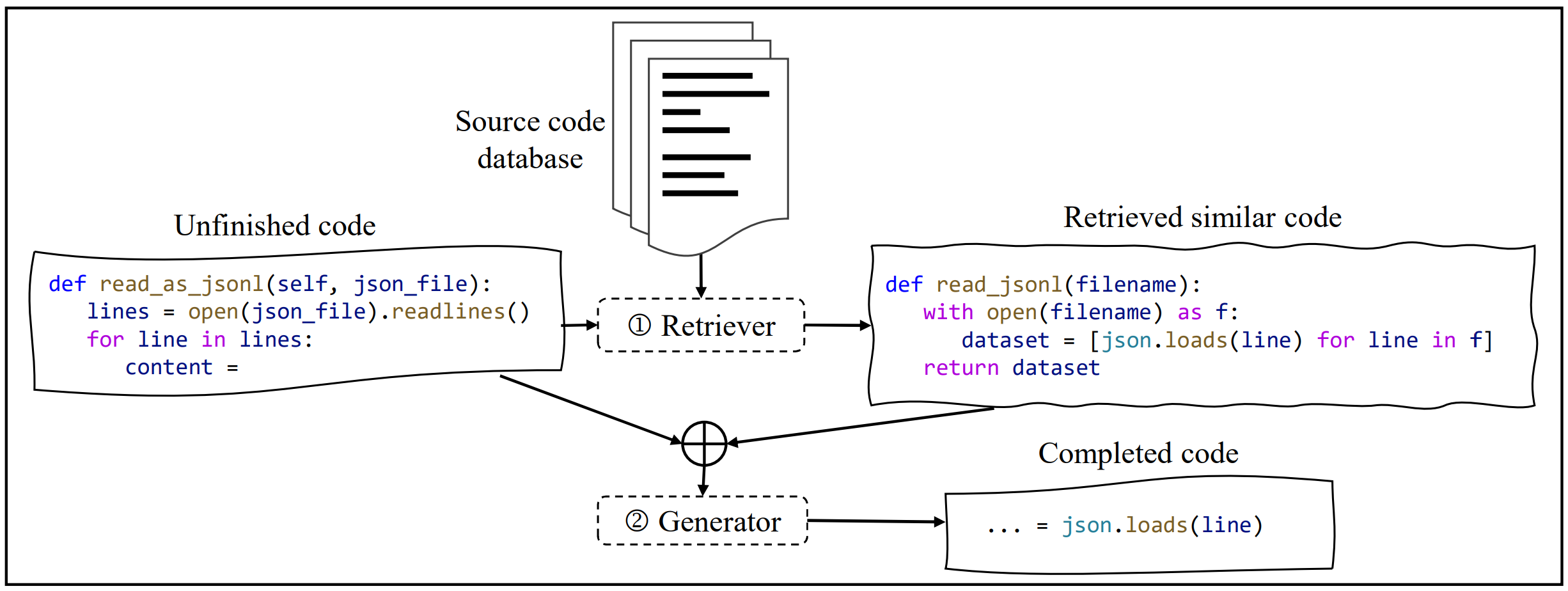}
\caption[ReACC Framework Illustration]{\textbf{ReACC Framework Illustration \cite{ReACC}.} This figure illustrates the ReACC framework, which retrieves similar code snippets from a database to use as external input for code completion.}
\label{fig:reacc}
\end{figure}

In contrast to general-purpose code completion frameworks like ReACC \cite{ReACC}, which leverages retrieval-augmented generation (RAG) without structural code analysis (see Figure \ref{fig:reacc}), our approach combines RAG with Abstract Syntax Tree (AST) analysis specifically to enable seamless integration of generated components into existing embedded system implementations.

\section{\textbf{Outlook}}
\label{sec:outlook}

\subsection{Conclusion}

The proposed method demonstrates significant potential for streamlining the code generation process, particularly in the domain of embedded systems. By automating the identification and generation of hardware-specific code, this approach minimizes the need for developers to manually look up microcontroller-specific details or select appropriate libraries. As a result, the code generation process becomes more adaptable across different microcontroller platforms, offering a flexible and efficient solution.

Based on expert feedback, future iterations of this prototype should focus on incorporating standardized hardware abstraction layers for defining hardware addresses. This enhancement would not only align the prototype more closely with industry standards but also effectively segregate hardware-specific code into distinct layers, addressing concerns regarding maintainability and scalability in more complex systems.

In light of these insights, future work will explore the following hypotheses to guide further development and improvements in the system.

\subsection{Future Work}

Further research is necessary to improve the system's capacity to handle increasingly complex and varied programming environments, as well as to integrate support for additional microcontroller architectures.

\paragraph{Hypothesis 1: Integration of Datasheet Information} Can the integration of datasheet information into the code generation process improve the applicability of the prototype across different microcontroller families?

\paragraph{Hypothesis 2: Enhancing Hardware Abstraction} Can the code generation process, with its AST-based implementation, be extended to function effectively across a broader range of microcontroller families?

\paragraph{Hypothesis 3: Similarity Calculation for Structural Variance and Output Valuation}
To better understand the quality and diversity of the generated code, future work could explore the application of more precise similarity measurement techniques. While functionality has been shown to remain stable across iterations, structural variance in the code could provide deeper insights into the adaptability and creativity of the system. A token-based approach, such as the one described in \cite{Zakeri2023}, offers a practical solution to quantify these structural differences by analyzing programming constructs at a granular level.
The methodology behind the CCFinder \cite{Kamiya2002}, which employ token-by-token comparisons for clone detection and similarity analysis, could be adapted for this purpose.

\section*{\textbf{Acknowledgment}}
Technical resources and project guidance were generously provided by AGSOTEC.

ChatGPT was utilized to generate sections of this work, including text and code.

\section*{\textbf{Author Contributions}}
\noindent 
\textbf{Sebastian Haug}: Conceptualization, Methodology, Software, Investigation, Writing - Original Draft, Visualization, Writing - Review \& Editing.\\
\textbf{Christoph Böhm}: Methodology,  Writing - Review \& Editing, Supervision.\\
\textbf{Daniel Mayer}: Conceptualization, Investigation, Resources, Writing - Review \& Editing, Supervision, Project Administration.

\printglossary[title={\textbf{Glossary}}]

\printbibliography[title=References, heading=bibintoc]

\end{document}

%% file: glossary.tex
\newglossaryentry{AI}{
    name=AI: Artificial Intelligence,
    description={The simulation of human intelligence}
}

\newglossaryentry{LLM}{
    name=LLM: Large Language Model,
    description={A type of artificial intelligence designed to understand and generate human-like text}
}

\newglossaryentry{API}{
    name=API: Application Programming Interface,
    description={A software interface for offering a service to other pieces of software}
}

\newglossaryentry{AST}{
    name=AST: Abstract Syntax Tree,
    description={A tree representation of the abstract syntactic structure of source code written in a programming language}
}

\newglossaryentry{ReACC}{
    name=ReACC: Retrieval-Augmented Code Completion,
    description={A framework that enhances code completion by leveraging external context from a large codebase}
}

\newglossaryentry{CMSIS}{
    name=CMSIS: Cortex Microcontroller Software Interface Standard,
    description={A hardware abstraction layer independent of vendor for the Cortex-M processor series}
}

\newglossaryentry{HAL}{
    name=HAL: Hardware Abstraction Layer,
    description={A layer of programming that allows a computer operating system to interact with a hardware device at an abstract level}
}

\newglossaryentry{RAG}{
    name=RAG: Retrieval-Augmented Generation,
    description={A process that enhances large language models by allowing them to respond to prompts using a specified set of documents}
}

\newglossaryentry{STM32F407}{
    name=STM32F407: High-\allowbreak performance Microcontroller,
    description={A microcontroller that offers the performance of the Cortex-M4 core}
}

\newglossaryentry{AURIX TC334}{
    name=AURIX TC334: 32-bit Microcontroller from Infineon,
    description={A microcontroller designed for automotive and industrial applications, featuring a 32-bit TriCore-\allowbreak architecture}
}

\newglossaryentry{LED}{
    name=LED: Light-Emitting Diode,
    description={A semiconductor light source that emits light when current flows through it}
}

\newglossaryentry{CortexM4}{
    name=Cortex-M4: 32-bit processor design from ARM,
    description={A 32-bit processor design optimized for real-time applications with low power consumption}
}

\newglossaryentry{GPIO}{
    name=GPIO: General-Purpose Input/Output Pin,
    description={A versatile pin on a microcontroller that can be configured as either an input or an output. As an \textbf{input}, it can read external signals such as button presses. As an \textbf{output}, it can control devices}
}

\newglossaryentry{Offset}{
    name=Offset: Relative distance of a specific register,
    description={The relative distance or position of a specific register or memory location within a hardware block, measured from a base address}
}

\newglossaryentry{Clock}{
    name=Clock: Synchronization signal for operations,
    description={The signal used to synchronize operations within a microcontroller or hardware system, ensuring consistent timing and execution of tasks}
}

\newglossaryentry{GPT4oMini}{
    name=GPT-4o Mini: Large Language Model from OpenAI,
    description={A compact variant of the GPT-4 language model designed for cost-efficient and versatile tasks}
}

\newglossaryentry{FAISS}{
    name=FAISS: Facebook AI Similarity Search,
    description={An open-source library for efficient similarity search and clustering of high-dimensional vectors}
}

%% file: literatur.bib
@article{Justyna2023,
    author = {J. Justyna},
    title = {Global Surge in Embedded Software Demand: Here is Why},
    journal = {DAC.digital},
    year = {2023},
    month = {12},
    url = {https://dac.digital/global-surge-in-embedded-software-demand-here-is-why/},
    note = {[Online]. Accessed: Aug. 29, 2024}
}

@misc{OutlineThenDetails,
    author = {Wenqing Zheng and S P Sharan and Ajay Kumar Jaiswal and Kevin Wang and Yihan Xi and Dejia Xu and Zhangyang Wang},
    title = {Outline, Then Details: Syntactically Guided Coarse-\allowbreak To-\allowbreak Fine Code Generation},
    year = {2023},
    eprint={2305.00909},
    archivePrefix={arXiv},
    primaryClass={cs.CL},
    url = {https://arxiv.org/abs/2305.00909},
    note = {Accessed: 2024-10-01}
}

@article{SelfPlanningCodeGeneration,
    author = {Jiang, Xue and Dong, Yihong and Wang, Lecheng and Fang, Zheng and Shang, Qiwei and Li, Ge and Jin, Zhi and Jiao, Wenpin},
    title = {Self-Planning Code Generation with Large Language Models},
    year = {2024},
    issue_date = {September 2024},
    publisher = {Association for Computing Machinery},
    address = {New York, NY, USA},
    volume = {33},
    number = {7},
    issn = {1049-331X},
    url = {https://doi.org/10.1145/3672456},
    doi = {10.1145/3672456},
    journal = {ACM Trans. Softw. Eng. Methodol.},
    month = {9},
    articleno = {182},
    numpages = {30}
}

@inproceedings{LLMRAGCodeGeneration,
    author = {Heiko Koziolek and Sten Grüner and Rhaban Hark and Virendra Ashiwal and Sofia Linsbauer and Nafise Eskandani},
    title = {LLM-\allowbreak based and Retrieval-\allowbreak Augmented Control Code Generation},
    booktitle = {LLM4Code 2024},
    organization = {ABB Research, Germany},
    month = {4},
    year = {2024}
}

@misc{ReACC,
    author = {Shuai Lu and Nan Duan and Hojae Han and Daya Guo and Seung-won Hwang and Alexey Svyatkovskiy},
    title = {ReACC: A Retrieval-\allowbreak Augmented Code Completion Framework},
    year = {2022},
    eprint={2203.07722},
    archivePrefix={arXiv},
    primaryClass={cs.CL},
    url = {https://arxiv.org/abs/2203.07722},
    note = {Accessed: 2024-10-01}
}

@inproceedings{RetrievalAugmentedCodeGeneration,
    author = {Parvez, Md. Rizwan and Ahmad, Wasi and Chakraborty, Saikat and Ray, Baishakhi and Chang, Kai-Wei},
    title = {Retrieval Augmented Code Generation and Summarization},
    booktitle = {Findings of the Association for Computational Linguistics: EMNLP 2021},
    year = {2021},
    month = {1},
    pages = {2719--2734},
    doi = {10.18653/v1/2021.findings-emnlp.232},
    organization = {Association for Computational Linguistics},
    address = {Punta Cana, Dominican Republic},
    url = {https://aclanthology.org/2021.findings-emnlp.232}
}

@misc{douze2024faisslibrary,
    title={The Faiss library},
    author={Matthijs Douze and Alexandr Guzhva and Chengqi Deng and Jeff Johnson and Gergely Szilvasy and Pierre-Emmanuel Mazaré and Maria Lomeli and Lucas Hosseini and Hervé Jégou},
    year={2024},
    eprint={2401.08281},
    archivePrefix={arXiv},
    primaryClass={cs.LG},
    url={https://arxiv.org/abs/2401.08281}
}

@misc{Urban2024,
    author = {Eric Urban},
    title = {Introduction to Prompt Engineering},
    year = {2024},
    url = {https://learn.microsoft.com/en-us/azure/ai-services/openai/concepts/prompt-engineering},
    note = {Accessed: 2024-03-29},
    howpublished = {\textit{Microsoft Learn}}
}

@misc{beningo2016apisvshals,
    author = {Jacob Beningo},
    title = {Embedded Basics -- API's vs HAL's},
    year = {2016},
    month = {4},
    url = {https://www.beningo.com/embedded-basics-apis-vs-hals/},
    note = {Accessed: 2024-08-21},
    organization = {Beningo\allowbreak Embedded\allowbreak Group},
    copyright = {© 2024 Beningo Embedded Group. All Rights Reserved.}
}

@misc{deepset2024,
    author = {deepset Cloud},
    title = {Prompt Engineering Guidelines},
    year = {2024},
    month = {4},
    url = {https://docs.cloud.deepset.ai/docs/prompt-engineering-guidelines},
    note = {Accessed: 2024-04-01}
}

@book{aho2006compilers,
    title={Compilers: Principles, Techniques, and Tools},
    author={Alfred V. Aho and Monica S. Lam and Ravi Sethi and Jeffrey D. Ullman},
    year={2006},
    edition={2nd},
    publisher={Pearson Education, Inc},
    address={Upper Saddle River, NJ},
    isbn={0-201-10088-6},
    lccn={QA76.76.C65 A37 1986},
    dewey={005.4/53 19},
    oclc={12285707}
}

@misc{STM32F4DISCOVERY,
    title = {STM32F4DISCOVERY},
    author = {{STMicroelectronics}},
    url = {https://www.st.com/en/evaluation-tools/stm32f4discovery.html},
    note = {(n.d.). [Online]. Accessed: 2024-08-29}
}

@book{zhu2015embedded,
    title={Embedded Systems with Arm Cortex-M Microcontrollers in Assembly Language and C},
    author={Yifeng Zhu},
    year={2015},
    edition={2nd},
    publisher={E-MAN PR LLC},
    address={Lulu, NC},
    isbn={0982692633},
    language={English}
}

@article{Zakeri2023,
    title = {A systematic literature review on source code similarity measurement and clone detection: Techniques, applications, and challenges},
    author = {Morteza Zakeri-Nasrabadi and Saeed Parsa and Mohammad Ramezani and Chanchal Roy and Masoud Ekhtiarzadeh},
    journal = {Journal of Systems and Software},
    volume = {199},
    pages = {111512},
    year = {2023},
    issn = {0164-1212},
    doi = {10.1016/j.jss.2023.111512},
    url = {https://doi.org/10.1016/j.jss.2023.111512},
    publisher = {Elsevier},
    note = {Available online 5 July 2023, Version of Record 17 July 2023}
}

@article{Kamiya2002,
    title = {CCFinder: A Multilinguistic Token-Based Code Clone Detection System for Large Scale Source Code},
    author = {Toshihiro Kamiya and Shinji Kusumoto and Katsuro Inoue},
    journal = {IEEE Transactions on Software Engineering},
    volume = {28},
    number = {7},
    pages = {654--670},
    year = {2002},
    month = {8},
    doi = {10.1109/TSE.2002.1019480},
    url = {https://doi.org/10.1109/TSE.2002.1019480}
}

@manual{RenodeFramework,
    title = {Renode: Open-Source Simulation Framework for Hardware Simulation and Testing},
    author = {{Antmicro}},
    year = {2024},
    note = {Available online at \url{https://renode.readthedocs.io/en/latest/}},
    url = {https://renode.readthedocs.io/en/latest/}
}
